\documentclass[twocolumn]{aastex7}
\usepackage{color}
\usepackage{amsmath}
\usepackage{amssymb}
\usepackage{mathtools}
\usepackage{upgreek}
\usepackage{enumitem}
\usepackage{gensymb}
\usepackage{graphicx}
\usepackage{bm}
\usepackage{multirow}
\usepackage{mathrsfs}

\usepackage{tabularx}
\usepackage[T1]{fontenc}
\usepackage{cleveref}

\usepackage{xcolor}

\crefname{subsection}{subsection}{subsections}

\shortauthors{}
\shorttitle{}

\begin{document} 

\title{Canceling Effects of Conjunctions Render Higher Order Mean Motion Resonances Weak}
i

\author[0009-0004-0844-173X]{Elizabeth K. Jones}
\affiliation{Department of Physics, Harvey Mudd College, Claremont, CA 91711, USA}
\email{lizzyjonesatx@gmail.com}

\author[0000-0002-1032-0783]{Samuel Hadden}
\affiliation{Canadian Institute for
Theoretical Astrophysics, 60 St George St Toronto, ON M5S 3H8, Canada}
\email{hadden@cita.utoronto.ca}

\author[]{Supakrai Teekamongkol}
\affiliation{Department of Physics, Harvey Mudd College, Claremont, CA 91711, USA}
\email{steekamongkol@hmc.edu}

\author[0000-0002-9908-8705]{Daniel Tamayo}
\affiliation{Department of Physics, Harvey Mudd College, Claremont, CA 91711, USA}
\email{dtamayo@hmc.edu}

\begin{abstract}
Mean motion resonances (MMRs) are a key phenomenon in orbital dynamics.
The traditional disturbing function expansion in celestial mechanics shows that, for coplanar systems at low eccentricities, $p$:$p-q$ MMRs exhibit a clear hierarchy of strengths, scaling as $e^q$, where $q$ is the order of the resonance.
This explains why first-order MMRs (e.g., 3:2 and 4:3) are important, while the infinite number of higher order integer ratios are not.
However, this relationship derived from a technical perturbation series expansion provides little physical intuition. 
In this paper, we provide a simple physical explanation of this result for closely spaced orbits in coplanar systems.
In this limit, interplanetary interactions are negligible except during close encounters at conjunction, where the planets impart a gravitational “kick” to each other's mean motion.
We show that while first-order MMRs involve a single conjunction before the configuration repeats, higher order MMRs involve multiple conjunctions per cycle, whose effects cancel out more precisely the higher the order of the resonance.
Starting from the effects of a single conjunction, we provide an alternate, physically motivated derivation of MMRs' $e^q$ strength scaling.
\end{abstract}

\section{Introduction} \label{sec:intro}

Configurations where two bodies' orbital periods form an integer ratio (e.g., 3:2 or 5:3) are known as mean motion resonances (MMRs) and play an outsized role in orbital dynamics. 
When planets are not in resonance, they overtake one another at effectively random azimuthal locations, and the associated gravitational perturbations largely average out. 
In resonance, however, these conjunctions recur at nearly fixed longitudes, allowing weak gravitational effects to accumulate coherently to large amplitudes.
These MMRs exhibit rich dynamics that can act to either stabilize or destabilize systems, playing a central role in shaping the architectures of planetary systems, rings, and disks \citep{Goldreich80, Murray99}. 

But why do only some MMRs matter?
Although any given period ratio is arbitrarily close to \textit{some} integer ratio (e.g., 1532:867), only a small subset of these ratios play a meaningful role in nature.
The reason is that, at small eccentricities $e$, the strength of a $p:(p-q)$ MMR scales as $e^q$ \citep{Murray99}, establishing a hierarchy in resonance strengths. First-order ($q=1$) MMRs are strongest (e.g., 2:1, 3:2 etc.), while higher-order resonances are progressively weaker at small eccentricities. 
This hierarchy is key to the existence of stable orbital regions; if all MMRs were comparably strong, their resonant regions would overlap and drive rapid chaos and instabilities \citep{Wisdom80, Hadden18}, precluding the existence of multiplanet systems. 

However, this only pushes the underlying question backward.
Why do the strengths of MMRs scale as $e^q$?
Traditionally, this scaling is derived from detailed perturbation theory expansions of the gravitational potential between two planets. 
Each MMR corresponds to a single cosine term in this ``disturbing function" expansion, whose amplitude scales as $e^q$ at leading order in eccentricity, as required by the d'Alembert rule \citep[e.g.,][]{Murray99}.
This is a powerful formalism, but it yields limited physical intuition.

\cite{Tamayo25} recently provided a simple qualitative explanation for this scaling in the case of coplanar systems.
The order of a resonance corresponds to the number of conjunctions that occur before the cycle repeats.
For example, in a 8:5 MMR, the inner planet must overtake the outer planet 3 times to complete 3 additional orbits during the cycle.
\cite{Tamayo25} argued that the reason higher order MMRs were weak was that the effects of such multiple conjunctions tend to cancel out.
However, their simple arguments could only show that the effects canceled out at first order in the eccentricity.

In this paper we perform a more careful analysis demonstrating that this intuition is correct, and that for higher order MMRs the cancellation extends to higher powers of the eccentricity. 
Indeed, we provide an alternate derivation of the traditional $e^q$ strength scaling of MMRs, starting from the effects of individual conjunctions. 
This provides a physical answer to a central principle in orbital dynamics, and yields a different approach that could be useful for various dynamical problems.

We begin in Sec. 2 by d efining our dynamical model and evaluating the effects of individual conjunctions building on work by \cite{Namouni96}. In Sec. 3, we apply this formalism to evaluate the strength of a first-order MMR. In Sec. 4 we extend the argument to higher-order MMRs, and show that that the kicks within a resonant cycle largely cancel. Finally, we show that the residual effect of such multiple conjunctions leads to the traditional $e^q$ MMR strength scaling.

\section{Dynamical Model} \label{chap:geometry}
\subsection{Hill Limit} \label{sec:Hill}
We begin by noting that at low eccentricities, most strong resonances correspond to close orbital spacings.
Even the 2:1, the most widely spaced first-order MMR, corresponds to a fractional separation $(a_2-a_1)/a_2 \approx 0.37$.
This motivates exploring the problem in the Hill limit of close orbital spacings where it is easier to make analytical progress \citep{Hill1878}.

The first advantage for tightly spaced orbits is that the interplanetary separation becomes so small at conjunction that one can ignore the gravitational interactions at all other times.
This allows the dynamics to be accurately modeled as a discrete map with impulsive gravitational “kicks’’ applied at each close approach, with the planets following unperturbed Keplerian orbits between conjunctions \citep{Duncan89}. 

The second major advantage is that in a close analogy to the two-body problem, the dynamics of two planets in the Hill limit separates into center-of-mass and relative degrees of freedom that can be independently solved \citep{Henon86}.
This makes it possible to identify a one-to-one mapping between the general problem of two massive planets on coplanar eccentric orbits and the much simpler circular restricted 3-body problem (CR3BP), with a massive planet on a fixed, interior circular orbit, and an exterior test particle on an eccentric orbit \citep{Hadden18}.
For a discussion of this equivalence and physical arguments for its existence, see \cite{Tamayo25}.

For mathematical and conceptual simplicity we therefore restrict ourselves to the CR3BP (Fig.\:\ref{fig:geometry}) where one only needs to consider a single planet mass $m$ (of the massive inner planet), and eccentricity $e$ and longitude of pericenter $\varpi$ (of the outer test particle). 
One can translate between this simpler problem and the general one of two massive planets on eccentric orbits using Table 1 of \cite{Tamayo25}.
This correspondence is exact in the Hill limit, and approximate at wider separations \citep[for more powerful and general results with traditional perturbative approaches, see][]{Hadden19}.

Finally, we note that an absolute eccentricity $e$ is not particularly informative in the closely spaced limit, as even small values might be orbit-crossing.
A more physically motivated quantity is 
\begin{align}
    \tilde{e} = \frac{e}{e_c} \label{e:etilde}
\end{align}
which measures how large the eccentricity is as a fraction of the crossing value $e_c$ at which the orbits would intersect one another.
\cite{Hadden18} show that expressing eccentricities in this way normalizes out the dependence of the dynamics on the orbital separation.
This implies that the numerical coefficients for MMRs of the same order (e.g., the 5:4 and 9:8), which are different in the traditional disturbing function \citep{Murray99}, become equal under this normalization for MMRs \citep{Hadden19}, resulting in a single universal coefficient for all MMRs of a given order at close separations \citep[see Sec.\:5.5 of ][]{Tamayo25}. 
At close orbital separations, the crossing eccentricity for a $p:p-q$ MMR is
\begin{align}
e_c \equiv \frac{\Delta a}{a} \approx \frac{2}{3} \frac{\Delta P}{P} \approx \frac{2q}{3p} \label{eq:ecross}.
\end{align}
where we have used Kepler's 3rd law (linearized for small differences between the semimajor axes $\Delta a$ and orbital periods $\Delta P$), and in the closely spaced limit it does not matter whether $a$ or $P$ in the denominator refer to the inner or outer planet.
We therefore use normalized eccentricities $\tilde{e}$ throughout the paper to simplify expressions and intuition.

\subsection{A Dynamical Model}\label{s:map_approach}

\begin{figure}[htb!]
    \centering
    \includegraphics[width=\linewidth]{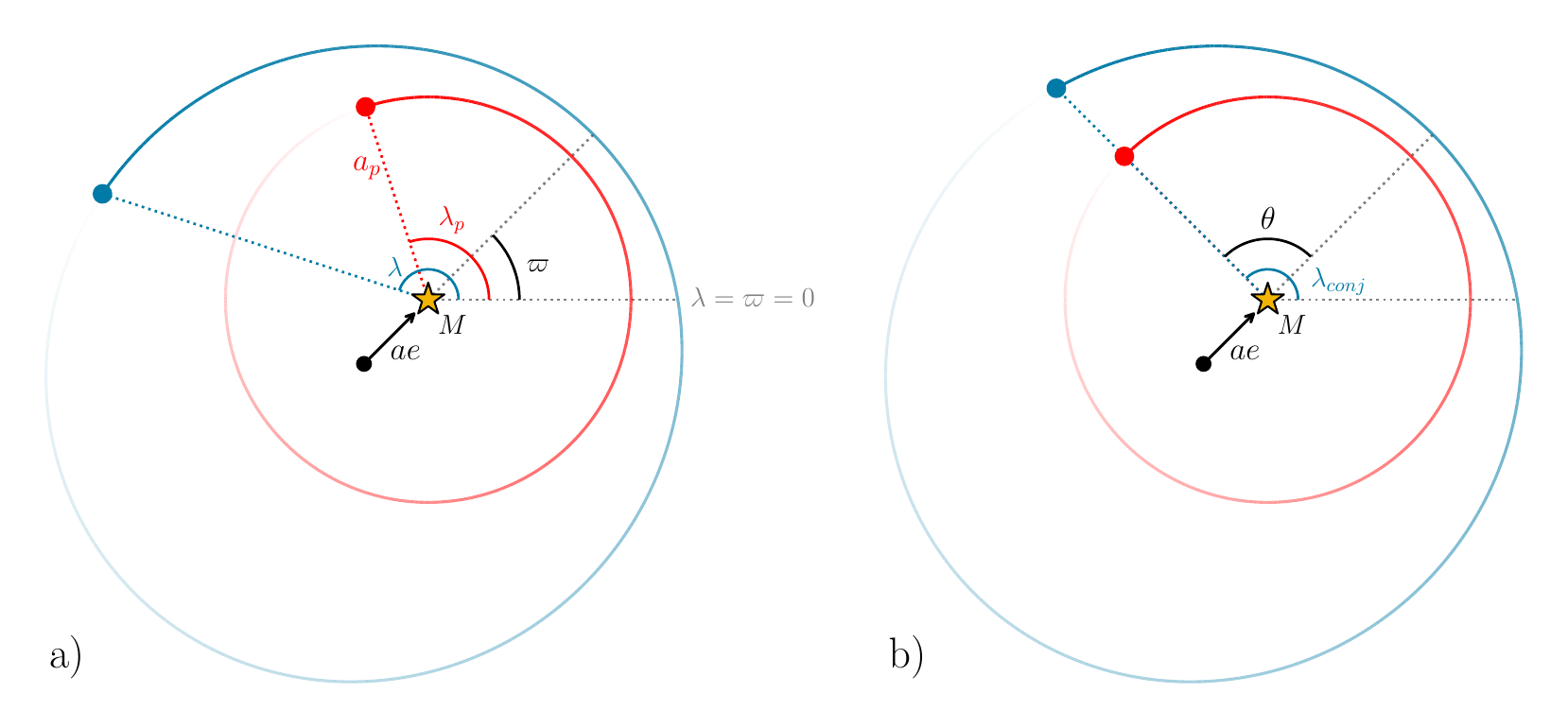}
    \caption{Illustration of the coplanar circular restricted three body problem, including the variables we use. Panel (a): the inner planet is on a circular orbit of radius $a_p$ at an angle $\lambda_p$. The outer test particle is on an elliptical orbit offset $ae$ from the center of the circle, where $e$ is the eccentricity of the ellipse. The orbit’s pericenter is located at $\varpi$ and the test particle is at the angular position $\lambda$. Panel (b): When the planets are at conjunction, they are both at the same longitude $\lambda_{\mathrm{conj}}$. The resonant angle is given by $\theta = \lambda_{\mathrm{conj}} - \varpi$. }
    \label{fig:geometry}
\end{figure}

We begin by noting that because the outer orbit is eccentric, the orbital separation varies with azimuth. 
Rather than referencing the longitude at which conjunction occurs $\lambda_{conj}$\footnote{Throughout the paper, we implicitly refer to the mean longitudes $\lambda$ as the actual locations of the planets. This is a useful intuition in the Hill limit where the eccentricities are necessarily small so the mean and true longitudes are approximately equal.
But we note that no approximations are made here, as only the mean longitudes enter the quantitative calculations below.
A conjunction is thus implicitly defined to correspond to when the \textit{mean} (rather than the true) longitudes align.} from an arbitrary direction (right panel of Fig.\:\ref{fig:geometry}), we choose to measure the conjunction angle $\theta$ from the point where the orbits are closest together\footnote{Always referencing conjunctions from the point of closest approach makes the dynamics symmetric in the alternative case where it's the inner orbit that is eccentric, as well as the general case of two eccentric orbits \citep[see discussion around Table 1 in][]{Tamayo25}.}. 
In our case with an outer eccentric orbit, $\lambda_{closest} = \varpi$ so we define the conjunction angle as $\theta \equiv \lambda_{conj} - \varpi$.

Let us take as a concrete example the 10:9 MMR, and as a first pass, begin by ignoring the interactions at conjunction.
If the planets begin at conjunction at some azimuthal angle $\theta$, then after 10 orbits of the inner planet, the outer test particle completes exactly 9 orbits, and the two planets have their next conjunction at exactly the same location. Turning on the gravitational interaction between the planets, such a resonance in principle allows the small perturbations at each conjunction to build up coherently, leading to much larger cumulative effects.

The complication is that these interactions slightly change the mean motion of the test particle, which perturbs the ratio slightly off the resonant value. 
If the test particle is now moving slightly too fast or slightly too slow (relative to the resonant value), the next conjunction will occur at nearly—--but not exactly—--the same location. 
One therefore has to consider the coupled evolution of how the conjunction location and mean motion change with time.

One advantage is that, because the individual kicks are small, the changes to the conjunction location $\theta$ are very small from one conjunction to the next.
One can therefore approximate this conjunction angle $\theta$ as a continuously varying quantity. 
For a $p:p-q$ MMR, the standard approach is to define the angle 

\begin{align}
    \theta \equiv \frac{p\lambda - (p-q)\lambda_p - q\varpi}{q} \stackrel{\mathrm{at\:conjunction}}{=} \lambda_{\rm{conj}} - \varpi, \label{eq:theta}
\end{align}
where $\varpi$ is the longitude of pericenter of the test particle \citep[e.g.,][]{Murray99}.
This makes $\theta$ well defined even between conjunctions, but one can see that at conjunction, where by definition $\lambda = \lambda_p = \lambda_\text{conj}$, $\theta$ corresponds to how far conjunction occurs from where the orbits are closest together (Fig.\:\ref{fig:geometry}).

\subsection{Pendulum Approximation}

We now make the pendulum approximation that the eccentricity and pericenter remain fixed, so called because it yields dynamics like that of a simple pendulum.

This approximation breaks down as $e$ approaches zero and the pericenter becomes undefined.
The more complicated second fundamental model of resonance captures this more subtle behavior \citep{Henrard83}, but even in these cases one can approximately match many of the same results through a simpler pendulum model as long as one takes the average (rather than the initial) eccentricity as the constant $e$ \citep{Tamayo25}.

To understand the dynamical behavior of our system, we aim to derive a differential equation for $\ddot{\theta}$, where overdots denote time derivatives.
While such an expression is given in Eq. 18 of \cite{Tamayo25}, we provide a simpler, self-contained derivation below.

We begin by taking a time derivative of Eq.~\ref{eq:theta} with our approximation that $\dot{\varpi} = 0$,
\begin{align}
    \dot{\theta} = \frac{p n - (p-q)n_p}{q} \label{eq:theta_dot},
\end{align}
where $n$ and $n_p$ are the mean motions of the test particle and the inner planet, respectively.
We also have that
\begin{align}
    \ddot{\theta} = \frac{d \dot{\theta}}{d n} \frac{d n}{d t} = \frac{p}{q}\frac{d n}{d t}, \label{eq:partials_theta_ddot}
\end{align}
where for the second equality we took a derivative of Eq.\:\ref{eq:theta_dot}, recalling that $n_p$ is constant.
To determine ${dn}/{dt}$, we remember that in the Hill limit, changes to the test particle's mean motion occur only at conjunction.
As long as we are in the typical perturbative limit where these kicks $\delta n$ to the mean motion are small, we can “smooth over’’ individual conjunctions, taking the continuous function $dn/dt$ to be the change in mean motion at conjunction divided by the time between successive conjunctions $\delta n / t_{conj}$.

The time between conjunctions $t_{conj}$ is the angular distance traveled by the inner planet between conjunctions divided by its constant angular rate $n_p$.
Since in one cycle of a $p:p-q$ MMR, the inner planet does $p$ orbits and must overtake its outer neighbor $q$ times during that period (Sec.\:\ref{sec:intro}), the inner planet must travel $2\pi p/q$ radians between each conjunction, yielding
\begin{align}
    \frac{dn}{dt} \approx \frac{\delta n}{t_{conj}} = \frac{qn_p}{2 \pi p} \delta n \label{eq:dn_dt}.
\end{align}
Substituting this into Eq. \ref{eq:partials_theta_ddot} yields
\begin{align}
    \frac{\ddot{\theta}}{n_p^2} = \frac{1}{2 \pi} \frac{\delta n}{n_p}, \label{eq:theta_ddot_eom}
\end{align}
where, because at close separations $n \approx n_p$, the right hand side can be interpreted as just depending on the fractional kick to the mean motion at conjunction.
Because this kick $\delta n$ is itself a function of $\theta$, this differential equation gives rise to interesting dynamics.

\subsection{Kicks at Conjunction} \label{chap:kick}

The key ingredient for understanding MMRs at close separations is thus the kick to the mean motion, and how it varies as a function of the azimuthal angle $\theta$ at which conjunction occurs. 
Given that the kick is a function of a periodic variable $\theta$, it can be expressed as a Fourier series.
The amplitudes of this Fourier series were worked out explicitly by \cite{Namouni96}.
We show in Appendix \ref{a:W_approx} that the fractional kick to the mean motion can be expressed as
\begin{align}
    \frac{\delta n}{n_p} \approx 2\pi \frac{\mu}{e_c^2} \sum_{j=1}^\infty j A_j^2 
    \tilde{e}^j \sin{(j\theta)} \label{eq:dn/n}, 
\end{align}
where $\mu = m/M_\star$ is the planet-star mass ratio, and the $A_j$ are order-unity numerical coefficients defined by \cite{Tamayo25} in such a way as to simplify the resulting expressions for MMR widths and libration frequencies (see their Table 2 for numerical values).

We see that the fractional change to the test particle's mean motion at conjunction scales with the planet-star mass ratio and fractional orbital separation as $\sim \mu / e_c^2$.
Additionally, for low eccentricities $\tilde{e} \ll 1$, Fourier terms diminish rapidly, with each subsequent term significantly smaller than the last.

A key point is that the size of these kicks is independent of what order MMR the planets are in, or whether they are in any MMR at all.
Instead, we will show that this Fourier expansion can explain the various well-known scalings for MMRs of different orders.

\begin{figure}[htb!]
    \centering
    \includegraphics[width=\linewidth]{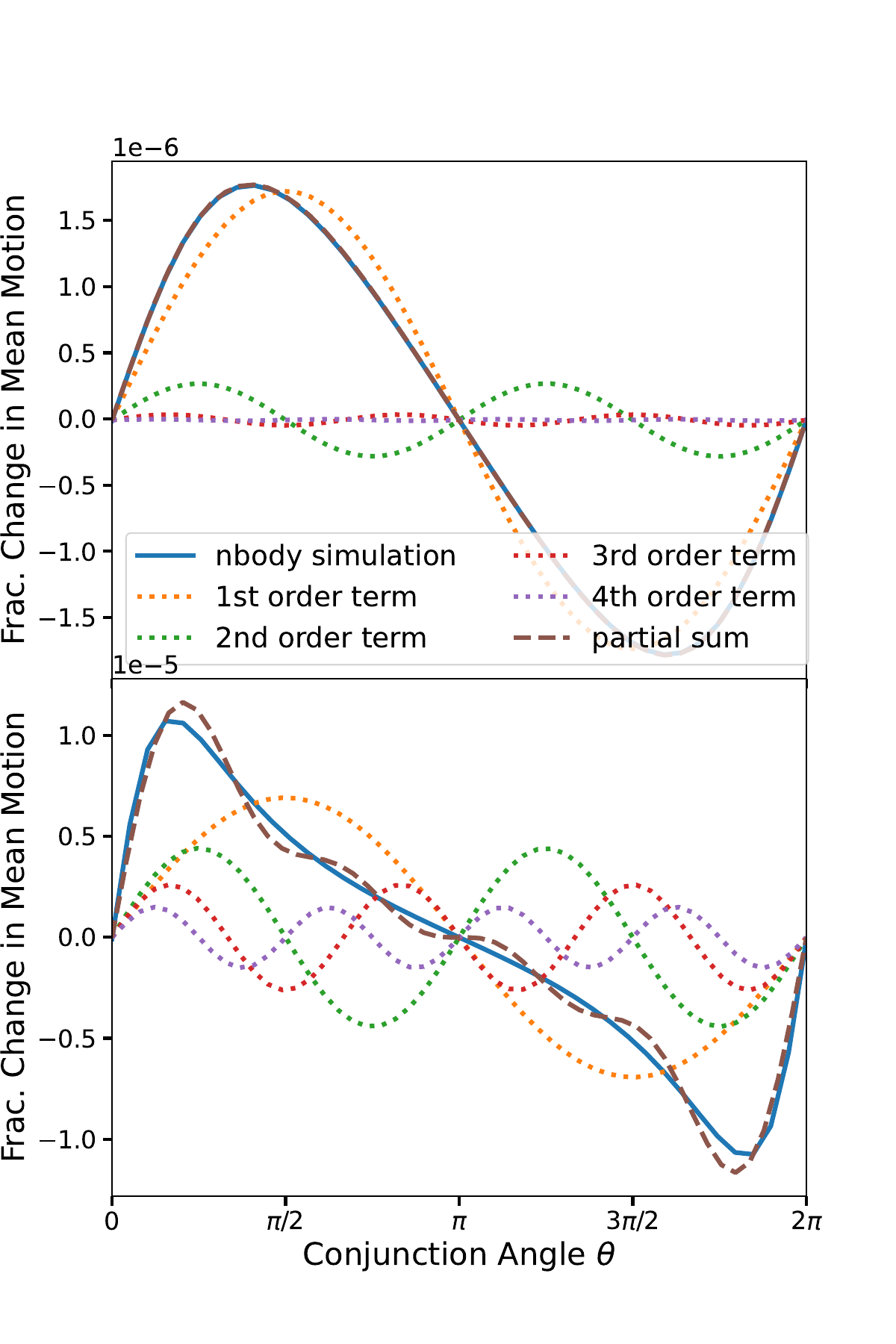}
    \caption{Fourier decomposition of the fractional change in mean motion $\delta n / n$ as a function of conjunction angle $\theta$ (Eq.\:\ref{eq:dn/n}), compared to results from direct $N$-body integrations. The N-body line (solid blue) simulates a suite of individual conjunctions at various angles $\theta$, and plots the resulting change to the mean motion.
    System parameters are given in the main text. The first four individual Fourier terms are plotted as dotted lines in orange, green, red, and purple, respectively; the dashed brown line shows their sum. Top panel: For small eccentricity ($\tilde{e} = 0.1$), the first-order term (orange) is a good approximation for the numerical result. Bottom panel: At higher eccentricity ($\tilde{e} = 0.4$), higher-order terms are comparable in magnitude to the first term and contribute significantly to the numerical result.}
    \label{fig:fourier_kick_decomp}
\end{figure}

As a numerical check in Fig.\:\ref{fig:fourier_kick_decomp}, we initialize a suite of numerical integrations in REBOUND with an eccentric test particle perturbed by a massive ($\mu=3.75 \times 10^{-10}$) planet on an interior, circular orbit.
The initial period is always chosen as $P_2/P_1 = 1.015$, which positions the bodies well within the closely spaced Hill limit, but outside any strong (low-order) MMRs.
Each integration is set up so that a single conjunction happens at a different $\theta$, and we numerically evaluate the resulting fractional change to the test particle's mean motion $\delta n/n$ due to the single gravitational encounter (blue solid line).
The top panel uses a test particle eccentricity $\tilde{e} = 0.1$, while the bottom panel uses a higher eccentricity $\tilde{e} = 0.4$. In the top panel, we see that the numerical results (solid blue line) closely match the sine curve shape of the first term in the Fourier series (orange dotted line), while the sum of the first four terms (dashed purple line) is visually indistinguishable from the integration. 
A single kick to mean motion thus contains contributions from all $\sin(j\theta)$ harmonics.

\section{First Order MMRs}

If we substitute the leading order term of the kick to the mean motion (Eq. \ref{eq:dn/n}) into our differential equation for $\theta$ (Eq. \ref{eq:theta_ddot_eom}), we obtain an expression of the form
\begin{align}
    \ddot{\theta} = +C_1 \sin{\theta},  \label{eq:pend_ddot}
\end{align}
which is the differential equation of a pendulum with its stable equilibrium $\theta=\pi$ away from where the orbits are closest together (i.e., where they are furthest apart)\footnote{Typically one defines $\theta'$ to be the deviation from the stable equilibrium of the pendulum, in which case $\ddot{\theta} = -C_1 \sin \theta$. But for reasons discussed by \cite{Tamayo25}, it is valuable to measure $\theta$ from where the orbits are closest together. One can go between the two definitions by plugging in $\theta = \theta' + \pi$.}. This corresponds to a Hamiltonian
\begin{align}
    H = \frac{\dot{\theta}^2}{2} + C_1 \cos \theta. \label{eq:Ham}
\end{align}
The ``strength'' of this first-order MMR $C_1$ is the coefficient of the $\cos{\theta}$ potential.
Plugging in our expressions above, we find
\begin{align}
    C_1 = A_1^2 \frac{\mu}{e_c^2} \tilde{e},
\end{align}
where the numerical coefficient $A_1 \approx 0.845$ \citep{Tamayo25}.
 
In the traditional celestial mechanics approach, the cosine term in Eq.\:\ref{eq:Ham} would be one of an infinite number of cosine terms in the disturbing function expansion, where one would average over all the other rapidly varying terms and isolate the single resonant term \citep[e.g.,][]{Murray99}. 

We see that the strength $C_1$ scales linearly with the eccentricity as expected for a first-order MMR.
However, while in the traditional approach there is a different numerical coefficient for each $p$:$p-1$ MMR \citep[see Appendix of][]{Murray99}, we see that in the compact limit, by expressing the eccentricity as a fraction of the orbit-crossing value, we instead obtain an order-unity coefficient $A_1$ that is universal to all first-order MMRs \citep{Hadden18, Tamayo25}.

This result is also physically plausible.
As discussed in detail by \cite{Tamayo25}, the outer planet gains energy as it approaches conjunction, and loses energy post-conjunction. 
If the orbits were circular, these exchanges would largely cancel.
But an eccentric orbit leads to an asymmetry in the separation between the planets and the strength of their interaction pre and post conjunction, leading to a net energy exchange \citep[see also][]{Peale76, Greenberg77, Peale86, Murray99}.
Thus, even if the effect might in general depend in a complicated way on the eccentricity, expanding around $\tilde{e}=0$ for small eccentricities, it makes sense that the strength would scale linearly in $\tilde{e}$ to leading order.

But this sets up the key question addressed by this paper.
Why do only the strengths of first-order MMRs scale linearly in $\tilde{e}$?
If the dynamics at close separations is dictated only by conjunctions, and each of these conjunctions lead to changes in the mean motion that scale linearly in the eccentricity (Eq.\:\ref{eq:dn/n}), why wouldn't MMRs of all orders share strengths of $\mathcal{O}(\tilde{e})$?

\section{Higher Order MMRs} \label{chap:higher}

The key difference is that, while first-order MMRs involve a single conjunction per resonant cycle, higher-order resonances feature multiple encounters, whose combined effects determine their overall strength. In this section, we extend our analysis to higher order MMRs and show how the geometry of these multiple kicks leads naturally to the well-known $e^q$ scaling of resonance strength. This approach provides a simple and intuitive geometric explanation for this central result in celestial mechanics.

We now wish to consider a $q$th order $p$:$p\!-\!q$ MMR. We define a `cycle' as the time the inner planet takes to complete $p$ orbits (and the outer planet completes $p\!-\!q$ orbits), after which the configuration repeats. After one such cycle, if we initially ignore the interplanetary interactions, the planets return to conjunction at the same azimuthal angle $\theta$, since both have completed an integer number of orbits.

Within each cycle, however, additional conjunctions now occur. Because the inner planet completes $q$ more orbits than the outer planet during a cycle, it must overtake the outer planet $q$ times. In other words, there are $q$ total conjunctions per cycle, which in general occur at different longitudes around the orbit. We investigate the specific locations of these conjunctions in Sec. \ref{s:kick_locs}.

In the regime where perturbation theory is useful, the dynamical evolution builds up over many cycles. 
To make analytical progress, we therefore assume that the mean longitudes advance at their unperturbed rates, and that the other orbital elements remain fixed over a  $p$:$p\!-\!q$ cycle in order to calculate the \textit{net} change to the test particle's mean motion $\Big(\frac{\delta n}{n} \Big)_{\mathrm{cycle}}$ from the $q$ conjunctions.
In essence, we move to considering a discrete mapping from one $p$:$p\!-\!q$ cycle to the next, rather than from each individual conjunction to the next. 
The same logic applied to obtain Eq.\:\ref{eq:theta_ddot_eom} then yields

\begin{align}
    \frac{\ddot{\theta}}{n^2} = \frac{1}{2\pi q} \Big(\frac{\delta n}{n} \Big)_{\mathrm{cycle}}, \label{e:thetaddot_cycle}
\end{align}
where
\begin{align}
    \Big(\frac{\delta n}{n} \Big)_{\mathrm{cycle}} = \sum_{k=0}^{q-1} \frac{\delta n}{n}\Big(\theta_k\Big).
\end{align}
and $\theta_k$ is the angle at which the $k$th conjunction in the cycle occurs. The additional factor of $\frac{1}{q}$ included in Eq. \ref{e:thetaddot_cycle} accounts for the fact that the net kick from the cycle is spread over a time that now spans $q$ conjunctions, so the denominator is a factor of $q$ longer in Eq.\:\ref{eq:dn_dt}.

Substituting in the expression for a single kick from Eq. \ref{eq:dn/n} yields
\begin{align}
    \left( \frac{\delta n}{n} \right)_{\mathrm{cycle}} = \sum_{k=0}^{q-1} \sum_{j=1}^{\infty} B_j \sin\left[j\theta_k\right] \label{e:cycle_kick1}
\end{align}
where
\begin{align}
B_j = 2\pi \frac{\mu}{e_c^2} j A_j^2 \tilde{e}^j. \label{eq:Bm}
\end{align}
We thus obtain a sum over $q$ conjunctions, each of which is an infinite sum over Fourier modes. 
Given that we assume $\tilde{e}$ is small, the $B_j$ shrink rapidly with increasing $j$.
This hierarchy makes it valuable to interchange the order of summation and exploit the fact that we are approximating the $B_j$ as equal for each conjunction within the cycle,

\begin{align}
    \left( \frac{\delta n}{n} \right)_{\mathrm{cycle}} = \sum_{j=1}^{\infty} B_j \Bigg(\sum_{k=0}^{q-1} \sin\left[j\theta_k\right]\Bigg). \label{e:cycle_kick}
\end{align}

This reorganizes the outer sum in terms of $B_j \propto \tilde{e}^j$ so that we can easily identify the leading order terms in the series. The inner sum is reduced to a geometric factor that only depends on the $q$ locations of conjunction $\theta_k$ within a cycle.

\subsection{Conjunction Locations Are Evenly Spaced}\label{s:kick_locs}

\begin{figure*}[htb!]
    \centering
    \includegraphics[width=\linewidth]{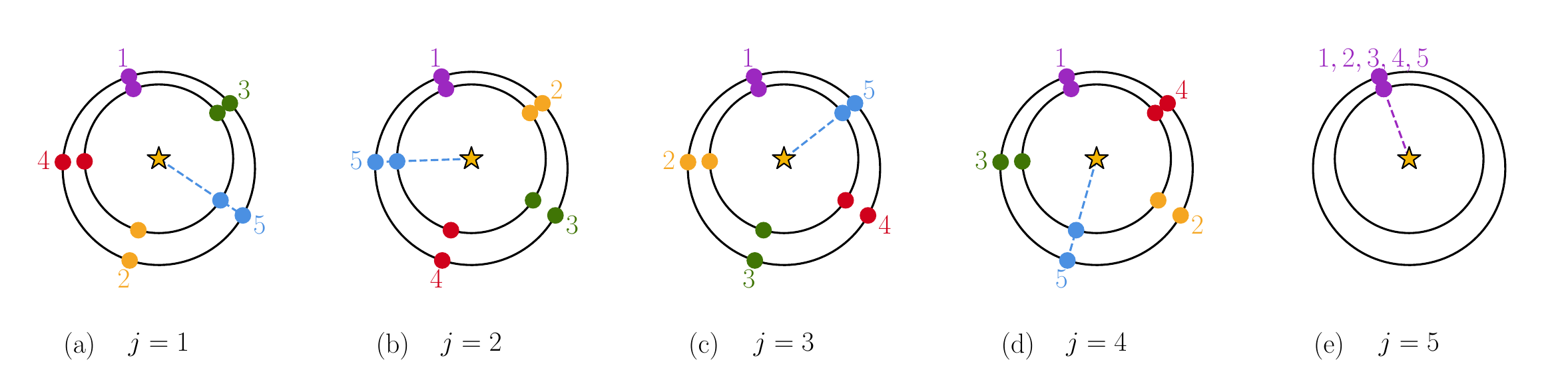}
    \caption{Angular locations of the conjunction angles multiplied by Fourier mode number $j$, shown for the 27:22 mean motion resonance. In all panels, the five original conjunction angles $\theta_k$ (for $k = 0$ to 4) are plotted around the orbit, with the ordering of each $j \theta_k$ indicated by color and label. Panel (a): For $j = 1$, the configuration of angles is equally spaced around the orbit, with a separation of $\Delta \theta = \frac{2 \pi p}{q}$ between subsequent conjunctions. Panels (b–d): For $j = 2, 3, 4$, the angular locations $j \theta_k$ remain evenly spaced around the orbit, but their order is permuted within the cycle, as shown by the changing sequence of colors. Panel (e): For $j = q = 5$, all $j\theta_k$ coincide at the angle $\theta$. }
    \label{fig:kick_locs_higher_m}
\end{figure*}

To evaluate Eq. \ref{e:cycle_kick} and determine the net effect of the $q$ kicks occurring during a cycle, we must identify the azimuthal locations $\theta_k$ of each conjunction along with their integer multiples $j \theta_k$. We aim to show that the assumption of constant Keplerian motion on nearly circular orbits always results in conjunctions with azimuthal locations that are evenly spaced around the orbit. 
This is not immediately obvious since the time between close approaches is longer than an orbital period and the conjunction locations jump around the unit circle (see the labels in panel (a) of Fig.\:\ref{fig:kick_locs_higher_m}).

To make analytical progress, we approximate the mean motion of the test particle as equal to the exact resonant value $n_{res}$ and constant over one resonant cycle.
While a more careful treatment would add higher order corrections to our expressions, this is sufficient to obtain the leading order result.


Recall from Sec. \ref{chap:higher} that a cycle of a $p:p-q$ MMR corresponds to the interval in which the inner planet completes $p$ full orbits. Thus, over the course of one cycle, the inner planet travels a total angular distance of $2 \pi p$. Since there are $q$ conjunctions per cycle and these conjunctions are equally spaced in time, the angular separation between successive conjunctions is
\begin{align}
    \Delta\theta = \frac{2\pi p}{q} \label{e:delta_theta}.
\end{align}
It follows that the $k$th conjunction (starting from the reference angle $\theta_0$) occurs at
\begin{align}
    \theta_k = \theta_0 + k \Delta \theta_0
\end{align}
where $k$ ranges from 0 to $q-1$.

In general, the offset $k \Delta \theta$ is not an integer multiple of $2\pi$, so these conjunctions occur at various longitudes around the orbit. Only when $k=q$ (and the cycle starts over) does the angular displacement become $2 \pi p$, returning the system to its original configuration.




As a concrete example, consider a $27:22$ MMR, where there are $27-22 = 5$ conjunctions per cycle. From Eq. \ref{e:delta_theta}, the angular displacement between successive conjunctions is
\begin{align}
    \Delta \theta_{27:22} = \frac{27}{5} 2\pi,
\end{align}
corresponding to five full revolutions and a remainder of $4 \pi / 5$ rad .
We plot the resulting five conjunctions in panel (a) of Fig. \ref{fig:kick_locs_higher_m}, labeled by their index $q$ for the order that they occur.

The fact that we assume motion at constant Keplerian rates implies that if conjunction 0 happened somewhere else, the entire five-fold pattern would be rotated correspondingly.
Additionally, if we started from any of the other conjunctions instead, this would yield the same five-fold repeating pattern.
Which conjunction is labeled as the first is therefore arbitrary.

We can then easily show that the conjunctions must be equally spaced in azimuth by contradiction.
For example, imagine that the conjunctions were irregularly spaced.
As argued above, if we started from conjunction 1 instead of conjunction 0, we should still get the same (irregular) five-fold pattern.
But if instead we rotated the original pattern so that conjunction 0 lined up with conjunction 1, the five-fold pattern would look different.
Thus, the only way for the two to agree is if the five-fold pattern is evenly spaced, as we sought out to show.

We will later need to show that not only are the $\theta_k$ equally spaced, but also that the angles $j\theta_k$ are equally spaced on the circle, for integer $j$. 
The same argument applies, since this case simply corresponds to starting the pattern at a different angle ($j\theta_0$) and using a different constant spacing between conjunctions ($j\Delta \theta$). 
The only exception is when $j$ is a multiple of $q$, since in that case $j\Delta \theta = 2\pi p$ (Eq.\:\ref{e:delta_theta}), so each of the conjunctions maps exactly to itself.
See Fig. \ref{fig:kick_locs_higher_m} for a visual representation.



\subsection{Summing Kicks within a Cycle} \label{s:adding_kicks}

We are now in a position to compute the total change in mean motion over one cycle. 

First, it is helpful to isolate the contribution of a single mode $j$ to this total cycle kick (i.e., the inner sum in Eq. \ref{e:cycle_kick})
\begin{align}
    \left( \frac{\delta n}{n} \right)_{\mathrm{cycle},j} = \sum_{k=0}^{q-1} B_j \sin\left(j\theta_k\right). \label{e:kick_for_mode} 
\end{align}

\begin{figure}[bht!]
    \centering
    \includegraphics[width=\linewidth]{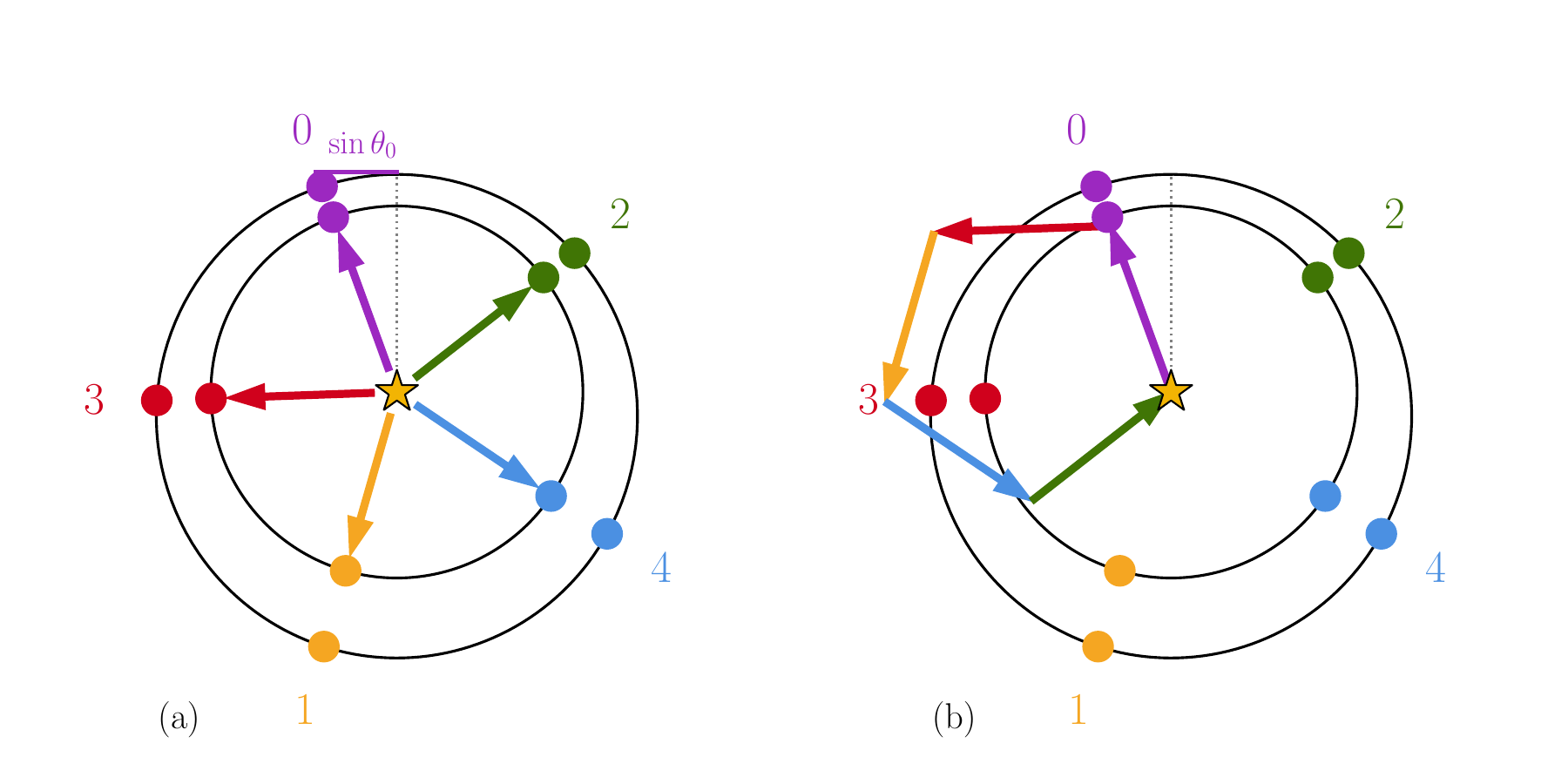}
    \caption{Geometric interpretation of the Fourier sum over kicks in a mean motion resonance cycle, shown for the case where the Fourier mode number $j$ is not divisible by $q$ (specifically, the 27:22 MMR with $q=5$). Panel (a): Each arrow represents a complex vector with angle $j\theta_k$, corresponding to one term in the sum of Eq.~\ref{e:kick_for_mode}. The five vectors are spaced uniformly around the unit circle by $2\pi/q$ and point in directions determined by the angles $j\theta_k$. Panel (b): Adding the vectors head-to-tail yields a closed regular polygon. As a result, their vertical components cancel exactly and the total contribution to the mean motion kick from this mode vanishes. This configuration corresponds to a sum over the $q$th roots of unity, which always vanishes when $j$ is not a multiple of $q$ \citep{rootsofunity}.}
    \label{fig:roots_of_unity}
\end{figure}

A valuable geometric interpretation for evaluating this sum is to imagine each arrow in Fig.\:\ref{fig:roots_of_unity} as a vector of magnitude $B_j$, pointing in the direction given by $\theta_k$.
Equation \ref{e:kick_for_mode} then says that the total fractional kick is given by the sum of the $x$-components of these vectors (see Fig.\:\ref{fig:roots_of_unity}).

If rather than directly summing the $x$-components of the vectors, we instead imagine first summing the vectors head-to-tail and then taking the $x$-component, the result becomes simple (Fig.\:\ref{fig:roots_of_unity}). 

In the case where the conjunction locations are equally spaced on the circle (when $j$ is not a multiple of $q$, i.e. panels a-d of Fig. \ref{fig:kick_locs_higher_m}), the vectors form the sides of a regular polygon that returns back to the origin.
Because the vector sum vanishes, the fractional kick to the mean motion cancels (Eq.\:\ref{e:kick_for_mode}).

Only when $j$ is a multiple of $q$ (panel e of Fig.\:\ref{fig:kick_locs_higher_m}) is there a net change to the mean motion, in which case the sum is simply
\begin{align}
    \left( \frac{\delta n}{n} \right)_{\mathrm{cycle},\, j{\mathrm{\, is \, multiple \, of \,}q}} 
    &= q B_j \sin{(j\theta)} \label{e:dn_m}
\end{align}

This geometric reasoning shows that only a subset of Fourier modes contribute to the cumulative effect of multiple conjunctions in a cycle: all modes with $j$ divisible by $q$ align constructively, while all other modes cancel out due to symmetry. 
This cancellation can also be seen as a consequence of the vanishing sum of the roots of unity, i.e., the complex roots of the polynomial $z^q - 1$. The $q$ distinct roots form the vertices of a regular polygon in the complex plane, as in the example shown in Fig.\:\ref{fig:roots_of_unity}.

We can now combine these results to derive a general expression for the net kick per cycle and recover the familiar eccentricity scaling of MMRs.

\subsection{Recovering the $e^q$ Scaling of MMR Strength}

Given that only values of $j$ that are multiples of $q$ survive, the leading order term corresponds to $j=q$. Plugging Eq.\:\ref{eq:Bm} into Eq.\:\ref{e:dn_m} we obtain
\begin{align}
    \frac{\delta n}{n} \Big(\theta\Big)_{\mathrm{cycle}} 
    &\approx  2\pi \frac{\mu}{e_c^2} q A_q^2 \tilde{e}^q \sin{(q\theta)}.
\end{align}
We can then plug into Eq.\:\ref{e:thetaddot_cycle} and bring the differential equation into pendulum form by identifying the relevant angle as $\phi = q\theta$.\footnote{In the dynamics literature, $\phi$ is usually referred to as the ``resonant angle", and is the angle that shows up in a given term in the traditional disturbing function expansion. For a first-order MMR, the resonant angle is thus just the angle $\theta$ at which conjunction occurs, but in general the resonant angle $\phi = q\theta$.}
This yields
\begin{align}
    \frac{\ddot{\phi}}{n^2} = q\frac{\ddot{\theta}}{n^2} = C \sin \phi
\end{align}
with
\begin{align}
    C =  \mu \Big(\frac{q A_q}{e_c}\Big)^2\tilde{e}^q.
\end{align}
This recovers the sought result that, to leading order in eccentricity, the strength of a $p:p-q$ MMR scales as $e^q$. 
For the corresponding resonance widths and oscillation frequencies, see Sec.\:6.3 of \cite{Tamayo25}.

\section{Conclusion} \label{chap:conclusion}

This paper provides a physical explanation for why higher order MMRs are weak.
Each cycle in which a high-order MMR configuration repeats entails multiple encounters, and the corresponding gravitational kicks from those close encounters cancel each other out more and more precisely the higher the order of the resonance.

One can also derive precise quantitative results from this geometrical picture in the closely spaced (Hill) limit.
By building on Fourier expansions for the orbital changes caused by a close conjunction \citep{Namouni96}, we were able to reproduce the well-known result that the strength of a $q$th-order MMR scales as $e^q$ to leading order.

This simple geometric approach should be useful to a wide variety of dynamical problems involving closely spaced planets.

\begin{acknowledgements}
We would like to thank an anonymous reviewer for their careful reading of the manuscript and insightful comments. 
\end{acknowledgements}

\appendix

\section{Change to the Mean Motion at Conjunction} \label{a:W_approx}

In this section we obtain the change to the mean motion at conjunction from the expressions in \cite{Namouni96}. 
They define an action associated with the orbital separation
\begin{align}
    K = \frac{3}{8} \Bigg( \frac{a-a_p}{a_p\epsilon} \Bigg)^2 = \frac{3e_c^2}{8 \epsilon^2} ,
\end{align}
where $\epsilon = \left( \mu / 3 \right)^{1/3}$ so $a_p \epsilon$ is the planet's Hill radius.
Taking a logarithmic derivative, we have for small fractional changes in a single conjunction,
\begin{align}
    \frac{\delta K}{K} \approx 2 \frac{\delta (a-a_p)}{a-a_p} = 2 \Bigg(\frac{a}{a-a_p}\Bigg) \frac{\delta a}{a} = \frac{2}{e_c} \frac{\delta a}{a}
\end{align}
Similarly taking a logarithmic derivative of Kepler's third law and recalling $e_c \equiv (a-a_p)/a$,
\begin{align}
    \frac{\delta n}{n} \approx -\frac{3}{2} \frac{\delta a}{a} = -\frac{3}{4}e_c \frac{\delta K}{K} = -2 \frac{\epsilon^2}{e_c} \delta K. \label{eq:dKconversion}
\end{align}

\cite{Namouni96} show that the change $\delta K$ at conjunction (as well as changes to all the other orbital elements) can be obtained through derivatives of an effective potential $W$.
We note that we diverge from their notation by defining the conjunction angle from the location where the orbits are closest together rather than from where they are farthest apart. 
This causes the stable equilibrium of the resonance to always be at a resonant angle $\phi = q\theta = \pi$, rather than alternating between 0 and $\pi$ depending on whether the MMR is even or odd order \citep{Tamayo25}. 
With this definition,
\begin{align}
   \delta K = -\frac{\partial W}{\partial \theta}, \label{eq:dKEq} 
\end{align}
and the effective potential $W$ has Fourier expansion \citep{Namouni96}
\begin{align}
W = \sum_{j=-\infty}^\infty (-1)^j W_j \mathrm{{e}}^{i j\theta} = W_0 + 2\sum_{j=1}^\infty (-1)^j W_j \cos{(j \theta)}
\label{e:Wn_exp}
\end{align}
where the alternating $(-1)^j$ arises from our definition for $\theta$ differing by $\pi$ from \cite{Namouni96} and the final equality uses the fact that both the potential $W$ and the Fourier coefficients $W_j$ are real. 

Each Fourier coefficient $W_j$ is a function of the eccentricity and can itself be expanded in powers of $\tilde{e} = e/e_c$ \citep{Namouni96}.
In our notation\footnote{The full expression in \citet{Namouni96} additionally accounts for nonzero inclination, but we consider only coplanar systems. As a result, we always have $k=0$ in their $W_j^{p, k}$ coefficients.},
\begin{align}
    W_j = \frac{\epsilon}{e_c} \sum_{p=j}^{\infty} W_j^{p,0} \, \tilde{e}^p + \dots \label{e:Wn_full}
\end{align}
where the first non-vanishing term is always $p=j$, and $p$ always has the same parity as $j$.
In other words, the first few terms in the potential read
\begin{align}
    W &= \frac{\epsilon}{e_c} \left( 
    W_0^{0,0} +
    W_0^{2,0} \tilde{e}^2 +  W_0^{4,0} \tilde{e}^4 + \dots \right) \\ 
   &\quad - \frac{2 \epsilon}{e_c} \left( 
    W_1^{1,0} \tilde{e} +
    W_1^{3,0} \tilde{e}^3 + W_1^{5,0} \tilde{e}^5 + \dots \right) \cos{\theta} \nonumber \\ 
   &\quad + \frac{2 \epsilon}{e_c} \left( 
    W_2^{2,0} \tilde{e}^2 + W_2^{4,0} \tilde{e}^4 + W_2^{6,0} \tilde{e}^6 + \dots \right) \cos{2\theta} + \dots \nonumber
\end{align}

\citet{Namouni96} tabulates numerical values for the coefficients $W_j^{p,0}$, and these values show that these coefficients decrease rapidly with increasing $p$: in the $j=1$ and $j=2$ series, for example, terms beyond the leading order $p=j$ are smaller by about an order of magnitude. Since each successive term is also multiplied by a higher power of the small parameter $\tilde{e}^2$, their contributions diminish even further. 

This combined suppression suggests that each Fourier coefficient can be well approximated by its lowest-order term:
\begin{align}
    W_j \approx \frac{\epsilon}{e_c} W_j^{j, 0} \tilde{e}^j. \label{e:Wn_approx}
\end{align}

\begin{figure}[htb!]
    \centering
    \resizebox{.99\textwidth}{!}{\includegraphics{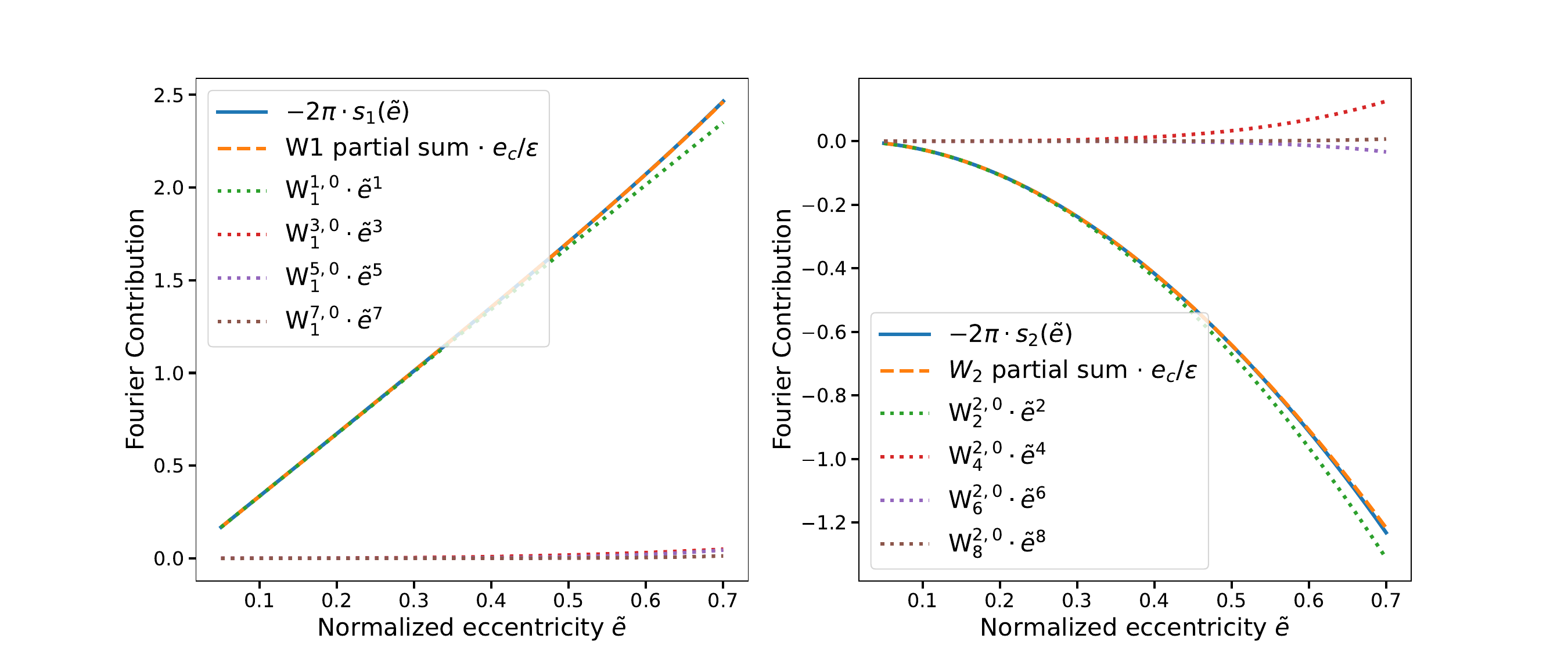}}
    \caption{Comparison between the Fourier coefficients of the effective potential computed via the Celmech function $\mathrm{s}_j(\tilde{e})$ (solid blue lines) and the truncated eccentricity expansions of the corresponding $W_j$ terms from Namouni's formalism \cite{Namouni96}. Left: The coefficient of $\cos(\theta)$ (i.e., $j=1$) is plotted as a function of normalized eccentricity $\tilde{e}$. The leading-order term $W_1^{1,0}\tilde{e}$ dominates, with higher-order terms $W_1^{3,0}\tilde{e}^3$, $W_1^{5,0}\tilde{e}^5$, and $W_1^{7,0}\tilde{e}^7$ contributing minimally. Their partial sum (dashed orange) closely tracks $-2\pi \cdot \mathrm{s}_1(\tilde{e})$. Right: Same comparison for $j=2$, corresponding to $\cos(2\theta)$. Again, the leading term $W_2^{2,0}\tilde{e}^2$ dominates, with higher-order corrections providing only small deviations. Even at $\tilde{e} = 0.7$, the leading-order approximations remain accurate to within a few percent, justifying their use throughout this work.}
    \label{fig:sk_W_comp}
\end{figure}

To verify this, we compare truncated series approximations (Eq.\:\ref{e:Wn_full}) to numerically evaluated integral expressions for the $W_j$ using the \texttt{celmech} package. 
The $W_j$ are constant multiples of the `sk' coefficients developed by \cite{Hadden18}
\begin{align}
    W_j = -2\pi \frac{\epsilon}{e_c} \mathrm{s}_j(\tilde{e}) 
\end{align}
Figure \ref{fig:sk_W_comp} shows this comparison for $j=1$ and $j=2$. Across the full range
$0 \le \tilde{e} \le 0.7$, the leading-order term $W_j^{j,0} \tilde{e}^{\,j}$ closely tracks the
corresponding $\mathrm{s}_j(\tilde{e})$ curve, with deviations of only a few percent even at the
highest eccentricities plotted. The eccentricities we consider in this paper are much smaller than this, placing our work firmly in the regime where the higher-order terms are negligible and the first-order approximation is sufficient.

We note that if we take only the leading order term in each $W_j$, $(-1)^j W_j^{j,0}$ has the advantage of always being negative, as can be easily verified from the numerical values in Appendix B of \cite{Namouni96}. 
We follow \citet{Tamayo25} in redefining the Fourier coefficients in terms of $A_j$, where
\begin{align}
A_j^2 = \frac{4}{3} \cdot \frac{|W_j^{j,0}|}{2\pi},
\end{align}
which simplifies expressions for resonance widths and oscillation frequencies \citep{Tamayo25}. 
Taking only the leading-order terms for each $W_j$ then yields a potential 
\begin{align}
    W \approx -3 \pi\frac{\epsilon}{e_c} \sum_{j=0}^\infty A_j^2 \tilde{e}^j \cos(j\theta).
\end{align}

Combining this expression with Eqs.\:\ref{eq:dKconversion} and \ref{eq:dKEq} finally yields the fractional kick to the motion we seek
\begin{align}
    \frac{\delta n}{n} = 2 \pi \frac{\mu}{e_c^2} \sum_{j=0}^\infty A_j^2 j \tilde{e}^j \sin(j\theta).
\end{align}

\bibliographystyle{aasjournal}
\bibliography{main}     
\end{document}